\documentclass[12pt]{article}
\usepackage{amsmath}
\usepackage{fullpage}
\usepackage{amsfonts}
\usepackage{graphicx}
\usepackage{epsfig}
\usepackage{setspace}
\usepackage{endfloat}
\newcommand{\ga}{\ensuremath{\left<}}
\newcommand{\dr}{\ensuremath{\right>}}
\newcommand{\moy}[1]{\ensuremath{\ga #1\dr}}
\newcommand{\cond}[2]{\ensuremath{\left.\ga #1\right|#2\dr}}
\newcommand{\EP}[1]{\ensuremath{\left(#1\right)}}
\newcommand{\F}[1]{\ensuremath{\overline{#1}}}
\newcommand{\T}[1]{\ensuremath{\widehat{#1}}}
\newcommand{\erreur}[2]{\ensuremath{\moy{\EP{#1-#2}^2}}}
\newcommand{\ds}{\ensuremath{\mbox{d}_s}}
\newcommand{\vs}{\ensuremath{\sigma_s^2}}
\newcommand{\fs}{\F{f(c)}}
\newcommand{\cb}{\F{c}}
\newcommand{\V}[1]{\vec{#1}}
\newcommand{\D}{\,\mbox{d}}
\newcommand{\dcb}{\epsilon_{\cb}}

\title{Optimal estimation for Large-Eddy Simulation of turbulence and application to the analysis of subgrid models}
\author{A. Moreau$^a$, O. Teytaud$^b$ and J.P. Bertoglio$^c$}
\date{\small $^a$ LASMEA, CNRS, Universit\'e Blaise Pascal, 24, avenue des
  Landais, 63177 Aubi\`ere, France.\\
$^b$ Equipe TAO (Inria), LRI, UMR CNRS 8623 (CNRS - Universit\'e Paris-Sud), Bat. 490 Universit\'e Paris-Sud 91405 Orsay, France.\\
$^c$ Laboratoire de M\'ecanique des Fluides et d'Acoustique, CNRS,
  Ecole centrale de Lyon, Université Lyon I, INSA de Lyon, 36 avenue
  Guy de Collongue, 69134 Ecully, France.}
\begin{document}
\maketitle

\abstract{The tools of optimal estimation are applied to the
study of subgrid models for Large-Eddy Simulation of turbulence.
The concept of optimal estimator is introduced and its properties
are analyzed in the context of applications to a priori tests of
subgrid models. Attention is focused on the Cook and Riley model
in the case of a scalar field in isotropic turbulence. Using DNS
data, the relevance of the $\beta$ assumption is estimated by
computing (i) generalized optimal estimators and (ii) the error
brought by this assumption alone. Optimal estimators are computed
for the subgrid variance using various sets of variables and
various techniques (histograms and neural networks). It is shown
that optimal estimators allow a thorough exploration of models.
Neural networks are proved to be relevant and very efficient in this
framework, and further usages are suggested.}

\section{Introduction}

The principle of Large Eddy Simulations is to solve the evolution
equations only for the large scales of a turbulent flow. Since the
large eddies do not contain all the information that would be
necessary to compute the future of a given flow\cite{langford},
the evolution equations for the large eddies are not closed. It is
then a generic problem in any type of LES, that the unknown terms
have to be approximated using known quantities, i.e. quantities
that can be computed directly using information associated with
the resolved field, or quantities estimated via additional subgrid
variables solution of auxiliary evolution equations (a subgrid
turbulent stress tensor \cite{schumann}, a subgrid probability
\cite{pope98}, or a subgrid spectrum \cite{bertoglio}...).

The aim of the present paper is to introduce the concept of {\em
optimal estimator} as a tool to estimate the minimal error that a
perfect subgrid model based on a given set of known large scale
quantities (the variables of the model) will generate (the
irreducible error). This optimal estimator strategy is considered
by the authors of the present paper as a very helpful process
in the field of subgrid modeling, since it provides a way of
assessing the relevance of the set of variables on which 
a subgrid model will be built, before having to specify the precise form of the
model.

The {\em optimal estimator} can be computed numerically if
the true subgrid term is known, that is to say using the results
of a Direct Numerical Simulation (DNS). It is therefore a concept
that is developed in the framework of what is usually referred to as "a priori"
test of subgrid models. The first sections of the paper
are devoted to presenting the method of building the optimal estimator using different
techniques. The classical technique relies on histograms, and a
new and promising technique based on neural networks is
introduced. It is pointed out in section 7 that neural
networks are indeed particularly relevant and efficient to build
the optimal estimator when the number of parameters to be included
in the subgrid model increases. Some classical results in optimal
estimation will also be exposed in the two following sections.

In the second part of the paper (sections 4 to 7), the interest of optimal estimators is illustrated
in the particular case of the subgrid modeling problem for a simple reaction term depending
on a single passive scalar in isotropic homogeneous turbulence. The procedure is used to assess the
performances of the popular Cook and Riley model\cite{cook},
which uses a $\beta$-distribution as a presumed form of the Filtered Density Function (FDF).

It is shown how it allows to distinguish between the various sources
of errors in the model. Once the error brought by each assumption
contained in the model has been estimated, it is easy to identify
at which level improvements could be made.

We will particularly compare the error associated with the choice of the $\beta$ distributions, with the one resulting from the use of a sub-model
to estimate the subgrid variance. We will address the problem of the relevancy of
the different variables which have been used in the literature\cite{cook,moin2000} to express the subgrid variance.

\section{Properties of optimal estimators}

The optimal estimator $\Omega$ for a quantity $\gamma$, from a set $\pi$ of quantities
(that we will call the {\em variables}) and a norm $||.||$, minimizes $||\Omega(\pi)-\gamma||$. Optimal
estimators are typically defined for the $L_2$-norm (quadratic error) and then
$\Omega$ minimizes $||\Omega(\pi)-\gamma||^2=\moy{(\Omega(\pi)-\gamma)^2}$ where
$\moy{.}$ is the statistical mean (also called the expectation).
The quantity $\moy{(\Omega(\pi)-\gamma)^2}$ is null if and only
if $\gamma$ is a deterministic function of $\pi$
, so that $\moy{(\Omega(\pi)-\gamma)^2}$ is generally not null.

A model is a function $g(\pi)$ which aims at approaching $\gamma$ (and
thus $\Omega(\pi)$) as closely as possible.
In a recent work\cite{langford} Langford and Moser
firmly asserted that the quadratic error is the relevant error to
consider in LES. This point of view is here adopted without further discussion and therefore
the quadratic error will be retained throughout the paper as the relevant criterium for assessing the quality of a subgrid model.
This section surveys and shows some properties of optimal
estimators in relation with the
quadratic error.

The quadratic error made by $g(\pi)$ when
estimating $\gamma$ is
\begin{equation}
E_g = \erreur{\gamma}{g(\pi)}.
\end{equation}

The quadratic error satisfies the following orthogonality relation (see the proof in appendix A)
\begin{equation}
\erreur{\gamma}{g(\pi)} = \erreur{\gamma}{\cond{\gamma}{\pi}} +
\erreur{\cond{\gamma}{\pi}}{g(\pi)}.
\label{e:orth}
\end{equation}

Since the last term in the RHS of equation \ref{e:orth} is the expectation of a
positive quantity, it is positive. Thus for any $g$
\begin{equation}\label{e:ineq}
\erreur{\gamma}{g(\pi)} \geq \erreur{\gamma}{\cond{\gamma}{\pi}}.
\end{equation}

This relation means that any subgrid model $g$, built on the set of variables $\pi$, will lead to quadratic errors larger
than the one made by the conditional expectation ${\cond{\gamma}{\pi}}$.

We will call the error made by the conditional expectation the
{\em irreducible error} made by estimating $\gamma$ using $\pi$,
since no model using $\pi$ as variables can
make a smaller error.
The last term in (\ref{e:orth}) can be written
\begin{equation}
\erreur{\cond{\gamma}{\pi}}{g(\pi)} =
\int
( \cond{\gamma}{\pi}-g(\pi) ) ^2\,p(\pi) \D\pi.
\end{equation}

It is equal to zero only if $g(\pi)=\cond{\gamma}{\pi}$. This means
that the conditional expectation is the unique best
model\cite{deutch60}. For the quadratic error, the optimal estimator
$\Omega(\pi)$ using $\pi$ as variables is thus the conditional
expectation $\cond{\gamma}{\pi}$.

Let $\pi'$ be a set of variables that can be computed using
$\pi$. The optimal estimator for $\pi'$ is thus implicitly a function of
$\pi$, so that equation (\ref{e:ineq}) becomes
\begin{equation}
\erreur{\gamma}{\cond{\gamma}{\pi'}} \geq \erreur{\gamma}{\cond{\gamma}{\pi}}.
\end{equation}
The irreducible error associated with $\pi'$ is then always greater than the
irreducible error associated with $\pi$.

The optimal estimators verify another property that is called the
``successive conditioning''
\begin{equation}\label{e:rec}
\cond{\gamma}{\cond{\gamma}{\pi}}=\cond{\gamma}{\pi}
\end{equation}
The proof of this property is given in appendix B. When
considering a given model, it is common to draw a cloud of
points with $\gamma$ on the $y$ axis and $g(\pi)$ on the $x$ axis\cite{cook,moin2000}. For
a given value $a$ of $g(\pi)$, it is possible to compute the mean of
$\gamma$ for all the points that are close to $a$. This can be done
for all the values of $a$ and drawn on the figure(i.e., {\em moving averages}). This is a
way of representing $\cond{\gamma}{g(\pi)=a}$, which is a function of $a$.
The successive conditioning of the optimal estimators means that if $g(\pi)$ is an optimal
estimator, then all the points computed as described should be on the
$y=x$ line.

Provided they can be computed, optimal estimators
(i) allow to know if a given model is
far from the optimal estimator by comparing the error it makes to
the irreducible error;
(ii) suggest ways of improving models just by representing the optimal estimators when this is
possible and (iii) allow to compare different sets of variables quantitatively, by computing the
irreducible error for each set.

\section{Practical computation of optimal estimators}

Now that the properties of optimal estimators
have been presented, we will turn to the practical computation of
optimal estimators using data: optimal estimation.
Optimal estimation {\em{from data}}
in non-parametric frameworks (i.e. without prior
knowledge of the function to be approximated) is a wide area of
research consisting in designing algorithms, termed {\em{learning algorithms}}, that use
data $(\pi_1,\gamma_1),\dots,(\pi_n,\gamma_n)$ to compute
approximations $f_n$ of the optimal $\Omega$, with some nice convergence
properties of $f_n$ to $\Omega$ as $n\to \infty$.

The main usual hypothesis is that the data are independent and
identically distributed. Various results of Universal Consistency
(UC), i.e. asymptotic convergence towards the optimal function in
$L^p$-norm, have been proved for various techniques; histogram-rules
(\cite{histo}), $k$-nearest neighbours \cite{knn}, neural networks
(\cite{white90}), gaussian support vector machines (\cite{svm});
various general results using VC-theory include wide families of
methods (\cite{generaluc}). There is no possible universal convergence
rate; a method is better or worse than another depending on the
distribution of the examples. However, various heuristics for choosing
between various learning-algorithms are well-known: support vector
machines are often efficient for generalizing from very small samples
or when relevant kernels can be defined, $k$-nearest neighbours only
need a metric, histograms are simple and interpretable, neural
networks do not work well in huge (non-sparse) dimensionality but can
deal with very large numbers of examples.

In consequence, two techniques have been chosen here, in the framework of $L^2$-norm
(quadratic error) using large DNS data.
The first one is the most intuitive
and is based on the fact that optimal estimators are conditional
expectations. This is the ``histogram technique''. The second one is based
on the fact that optimal estimators minimize the quadratic error. It uses neural networks.

First, a conditional expectation can be approximated by a {\bf
  piecewise-constant function} (a histogram).
The $\pi$ space (whose dimension is the number of variables of the
model) is discretized in small cells. Each data point (a value of
$\gamma$ and of $\pi$ that has been produced by a DNS)
belongs to a given cell of the $\pi$ space.
Let us consider
the piecewise-constant function which associates to each cell the mean of
$\gamma$ for all the data points belonging to the cell.
This function
is an approximation of the optimal estimator. When you have $\pi$, you
can take as an approximation of $\cond{\gamma}{\pi}$ the value of
the previous function for the cell corresponding to $\pi$.

The main difficulty is the choice of the size of the cell. If the size
is too big, the piecewise-constant function will obviously not be a good
approximation of the conditional expectation. If the size of the cell is too
small, too few points will be contained in each cell and the value
associated with each cell will not be reliable.

\begin{figure}
\centerline{\includegraphics[width=8cm]{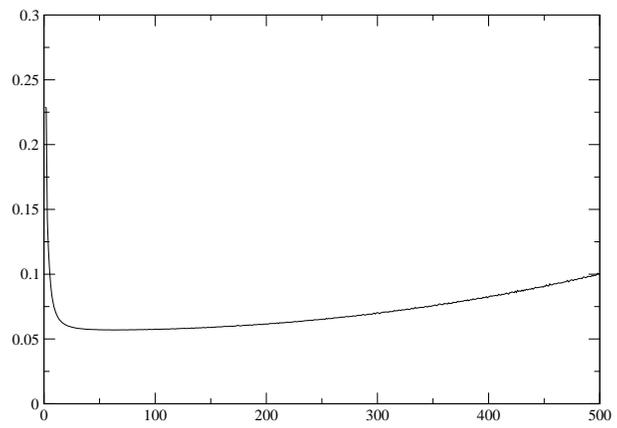}}
\caption{Typical generalization error versus the number of cells for
  each parameter.\label{fig:generalisation}}
\end{figure}

In order to overcome this difficulty, one has to divide the data into two
parts. The first one is used for the computation of the piecewise-constant
function. Then the error made by the piecewise-constant function when
estimating $\gamma$ for the {\underline second part} of the data will be computed.
This error is the {\em generalization error}. The relevant size for
the cells is the one for which the generalization error is minimized.
Figure \ref{fig:generalisation} shows the generalization error in
function of the number of cells for a given range.

\begin{figure}
\centerline{\includegraphics[width=8cm]{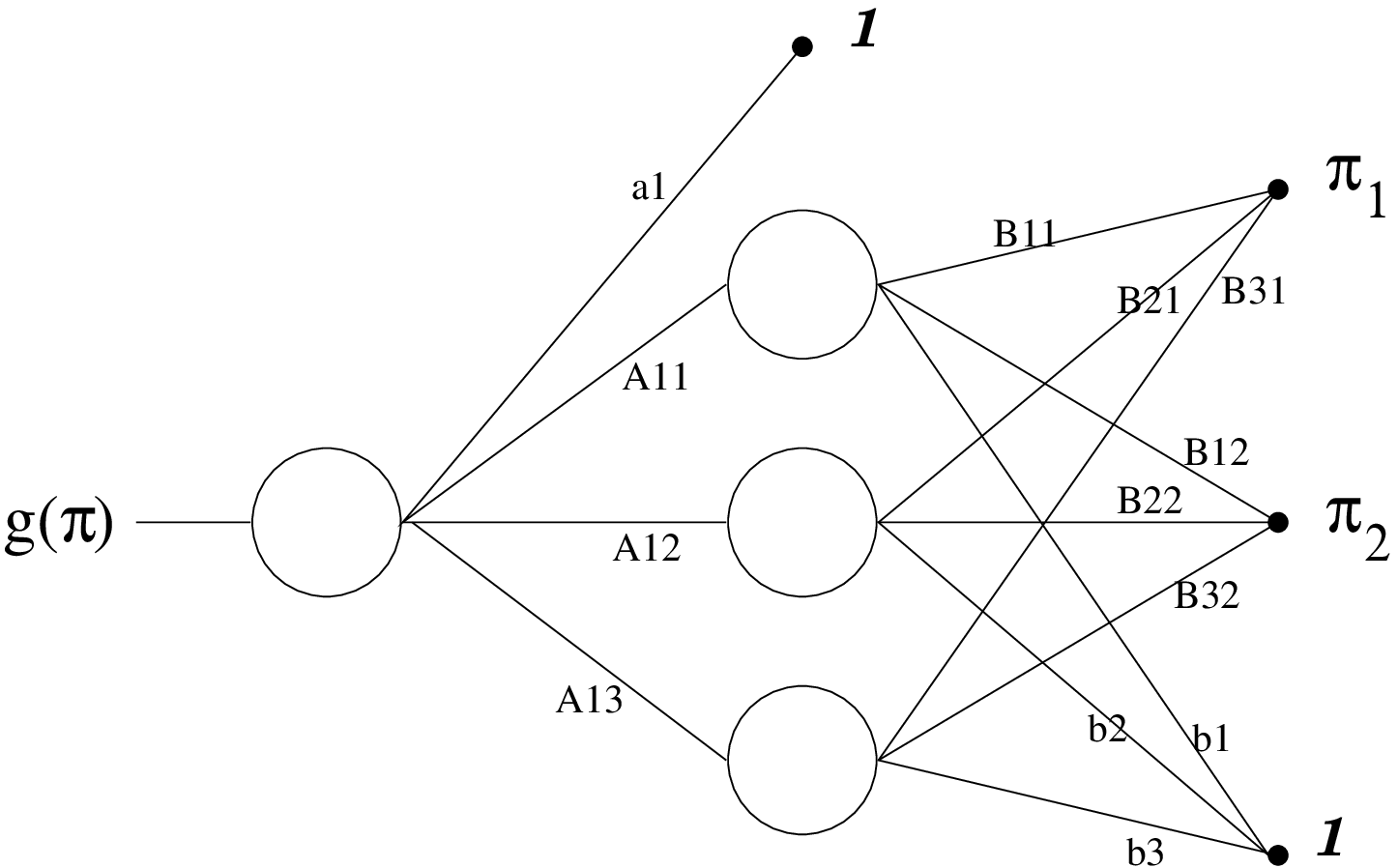}}
\caption{Representation of the neural network used to approximate the
  optimal estimators in the case where the number of neurons in the
  hidden layer is 3 and with two variables only.\label{fig:reseau}}
\end{figure}

The optimal estimators can be approached using {\bf neural networks} instead of piecewise-constant functions. A
neural network can be seen as a parametric function. For a perceptron with
a single hidden layer\cite{bishop}, this function can be written
\begin{equation}
g(\pi) = \sum_{j=1}^{N} A_{j} \, \mbox{tanh} \EP{\sum_{k=1}^{N_\pi} B_{jk}\,\pi_k+b_j}+a
\end{equation}
where $N$ is the number of neurons in the hidden layer, $N_\pi$ is the
number of variables in $\pi$ and $\pi_k$ is the $k^{th}$ parameter.
In the NN-terminology, the parameters $B_{jk}$ are called the weights of the first layer,
the $A_j$ are the weights of the second layer; $a$ and the $b_j$ are the thresholds.
By adjusting the weights of a neural network, it is
possible to approach almost any function. Formally, neural networks with one hidden layer of neurons have the universal approximation
property, i.e. they can approximate any measurable function for the
$L^2$-norm, and they have the
statistical consistency, i.e. this convergence occurs with probability one when
the network is trained from data if the number of neurons increases properly\cite{white90}.
A typical neural network is represented figure \ref{fig:reseau}.

As previoulsy, the data is split into two
parts. Using the first part, the neural network is
{\em trained}: the weights are adjusted so that the error made by
the neural network is minimized. This means that {\em the neural network is
a numerical approximation of the optimal estimator}. The learning is
made using a back-propagation algorithm\cite{rumelhart,bishop}.

Then, the {\em generalization error} is computed. The generalization error is the
quadratic error made by the neural network on the rest of the
data (on the data that have not been used for choosing the weights).
The number of hidden neurons is chosen in order to minimize this generalization error.

Neural networks allow to compute optimal estimators for a number of
variables which is greater than 3. Let us just stress that neural
networks are usual tools for pattern recognition, estimation of
conditional expectations, density estimation \cite{bishop}.
They have been applied in various areas of physics
(\cite{haykin,teytaudnn}), even in fluid mechanics\cite{lee97}. Other
forms of statistical learning tools derived from neural networks have
also been experimented, particularly Support Vector Machines
(\cite{mukherjee,ajjar}). We have chosen neural networks
because Support Vector Machines, at least in their most standard form,
do not use the mean square error, and therefore are not
conditional-expectation estimators, whereas standard neural networks
are\cite{white90}. However, less standard forms of Support Vector
Machines could also be used\cite{lssvm}; but SVM are much
slower than neural networks for large data sets such as the ones we will use further on.

The results obtained using neural networks are presented section \ref{s:nn}.

\section{Filtered Density Functions}

The case of a scalar field $c(\V{x})$ advected by turbulence is
now considered ($c(\V{x})$ representing for instance a temperature
or the concentration of a chemical species). The large scales of
the scalar field are denoted by $\cb(\V{x})$ defined as
\begin{equation}
\F{c}\EP{\vec{x}}=\frac{1}{h^3}\int_{D(\vec{x})}c(\vec{x'},t)~\mbox{d}\vec{x'}
\label{e:filtre}
\end{equation}
where $D(\vec{x})$ denotes a cube of edge-length $h$, centered in
$\vec{x}$. The $\F{.^{}}$ filter considered here is then a box
filter in the physical space. It is a positive filter: the
filtering can be written as a convolution $\cb=G*c$ of the scalar
field by a function
\begin{equation}
G(\V{x}) = \frac{1}{h^3} \prod_{i=1}^{3} \,
H\left(\frac{h}{2}-\left| x_i \right| \right)
\end{equation}
which is positive (see appendix C). Here, $H$ is the Heaviside
function. In the Fourier space, the filter corresponds to a
product of the Fourier transform of the scalar fluctuation by
\begin{equation}\label{e:transfert}
\widetilde{G}(\V{k})=\prod_{i=1}^{3} \mbox{sinc}
\left(\frac{1}{2}\,k_i\,h\right).
\end{equation}
In this article, the scalar field is bounded: $0\leq c(\V{x})
\leq 1$. As already pointed out\cite{cook}, the large eddies are
bounded in the same way only if $G\geq 0$ and $\int G =1$.

We now consider quantities that can be written as:
\begin{equation}
\F{f(c)}(\V{x})\label{fa}
\end{equation}
where $f$ is a function. It is not specified here, but $f(c)$ represents
a quantity that is important for the simulation and whose filtered
value requires closure: a (simple) chemical reaction term for
example. The quantity $\overline{f(c)}(\V{x})$ only depends on the Filtered
Density Function (FDF) which is defined\cite{pope} by
\begin{equation}
\ds(C,\V{x}) = \F{\delta\left(C -
c(\V{x'})\right)}(\V{x})\label{fdf}
\end{equation}
The link between the FDF and the quantity of interest is the
following relation\cite{pope}
\begin{equation}\label{eval}
\fs(\V{x}) = \int f(C) \, \ds(C,\V{x}) \D C
\end{equation}
which means that the knowledge of $\ds$ (which is sometimes called
the Subgrid PDF) allows to compute $\fs$ whatever $f$ is. The FDF
is not a statistical quantity: it is defined for a given
realization of the flow and for a given $\V{x}$. Now that this has
been underlined, the space dependence of the FDF will be omitted
in the following - as for $\fs$ or $\cb$.

The mean (on the cube $D(\V{x})$ and not in the statistical sense)
of the FDF is its first moment. It is simply equal to $\cb$
since
\begin{equation}
\int x \,\ds(x)\D x = \F{\int x \,\delta(x-c) \D x} =\cb.
\end{equation}
The variance of the FDF will be called the {\em subgrid variance}
\begin{equation}
\vs = \int (x-\cb)^2 \,\ds(x)\D x = \F{c^2}-\cb^2.
\end{equation}

For a given cube $D=D(\V{x})$, let us consider the proportion of the
cube which contains a scalar field bounded above by
$C$. It will be noted $V_D(C)$ and can be written
\begin{equation}
V_D(C) = \frac{1}{h^3} \int H(C-c(\V{x})) \D\V{x},
\end{equation}
where $H$ is the Heaviside function. We then have the following relation
\begin{eqnarray*}
\frac{\partial V_D}{\partial C} &=& \frac{1}{h^3} \int
\frac{\partial}{\partial C} H(C-c(\V{x})) \D\V{x}\\
&=&\frac{1}{h^3} \int \delta(C-c(\V{x})) \D\V{x} = \ds
\end{eqnarray*}
The FDF thus gives information about the distribution of the values of the scalar field
in the cube $D(\vec{x})$ since $V_D(C) = \int_{-\infty}^{C}
\ds(x)\D x$. Points can be chosen randomly in the cube
$D(\V{x})$. A histogram made using the values of the scalar at these
points will approach the FDF.
This gives a way of computing the FDF provided that the field is known
everywhere in the cube $D(\V{x})$.

We used data from a pseudo-spectral DNS with periodic boundaries. In such a simulation, the
evolution equation of the flow is solved in the Fourier
space. The fields of the velocity and of the scalar are then defined
by their first Fourier modes. For the scalar, this can be written
\begin{equation}\label{e:TFI}
c(x,y,z) =
\sum_{j=-\kappa}^{+\kappa}
\sum_{k=-\kappa}^{+\kappa}
\sum_{l=-\kappa}^{+\kappa}
a_{j,k,l} \, e^{i \omega_0 (j\,x+k\,y+l\,z)},
\end{equation}
where $\omega_0 = \frac{2\pi}{L}$ ($L$ being the size of the
simulation domain) where $\kappa$ is the number of modes retained and
the $a_{i,j,k}$ coefficients are the amplitudes of the Fourier modes.
The Fast Fourier Transform (FFT) computes the field at special points (grid nodes) because
(i) there is a mapping between the values of the fields at the grid nodes and
the amplitudes of the Fourier modes and (ii) equation (\ref{e:TFI})
can be factorized, so that the computation of the field at the grid
nodes is easier. But let us stress the fact that
the field is defined everywhere in the physical space by (\ref{e:TFI}). It can be
computed for an arbitrary $\V{x}$ by summing the contributions of the
different modes at this particular point.
This operation is of course
very costly compared to a FFT, but it is necessary. We have tried to
approximate the field between the nodes using a simple interpolation,
which requires less computation time, but the results are not satisfactory. The FDF
computed using interpolation substantially differs from the one
computed using the rigorous formula (\ref{e:TFI}).

The box filter (\ref{e:filtre}) cannot easily be computed in the
physical space. On the contrary, it is simple to perform in the
Fourier space, since the amplitudes of the Fourier modes just have
to be multiplied by a function given by (\ref{e:transfert}). This
is a rigorous method in the sense that the obtained field exacly
satisfies (\ref{e:filtre}) in the physical space, reflecting the
fact that the field is indeed implicitly defined everywhere in the
physical space when the Fourier modes are known.

\begin{figure}
\begin{center}
\centerline{\includegraphics[width=8cm]{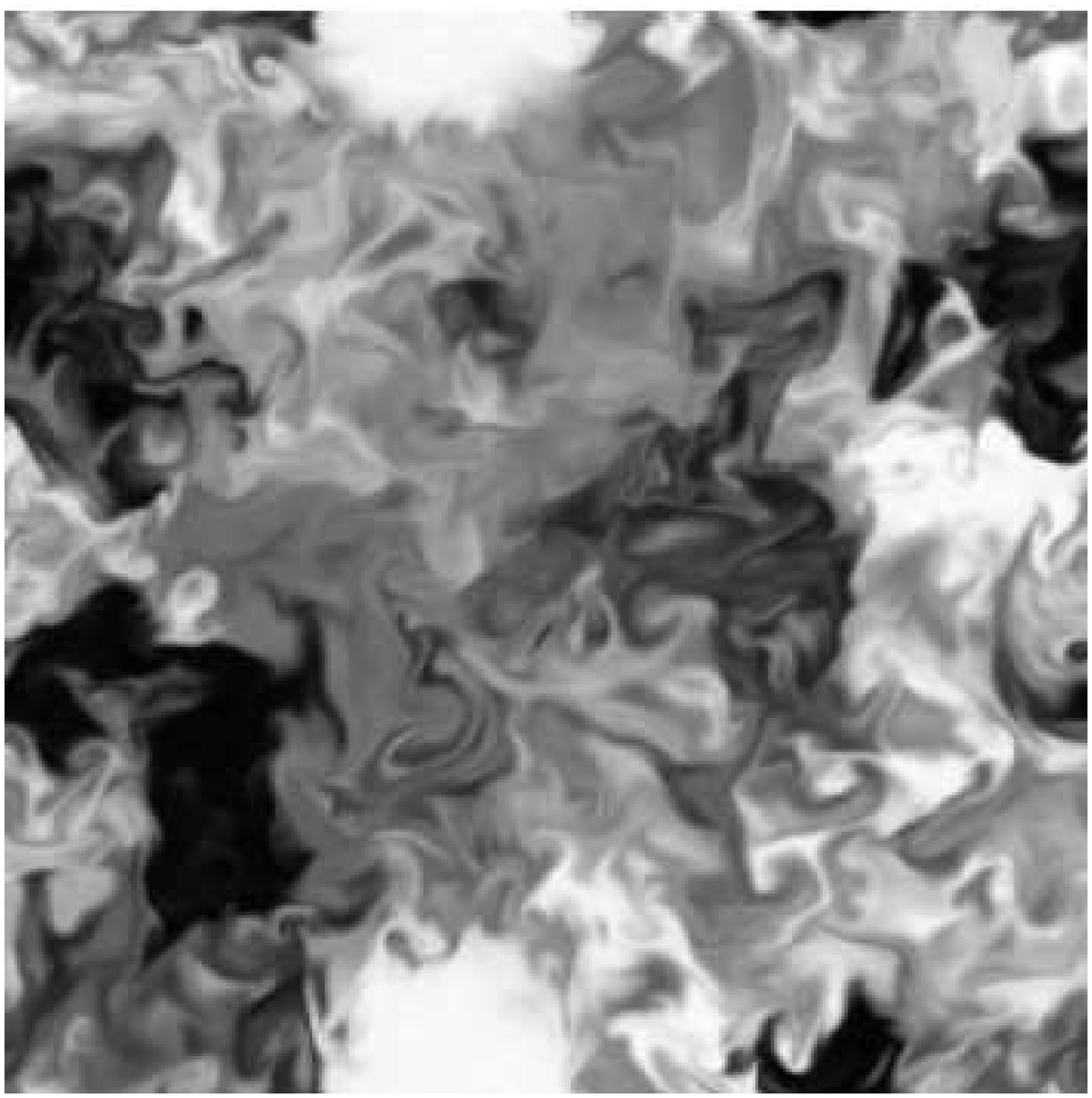}}
\caption{A view of the scalar field for a $256^3$ DNS. A recent injection can be seen, appearing
as a white homogeneous zone.\label{scalar}}
\end{center}
\end{figure}

The DNS we have performed use a particular injection method for
the scalar field. Periodically in time, large cubes are chosen
randomly in the simulation domain. ``Fresh'' scalar is injected in
these cubes: in half the cubes the field is put to zero and in
the other half it is put to 1. A view of the scalar field is shown
figure \ref{scalar}. It has to be stressed that the scalar
fluctuation always satisfies $0\leq c \leq 1$. The characteristics
of the simulations are detailed in \cite{elmo}.

\begin{figure}
\begin{center}
\centerline{\includegraphics[width=8cm]{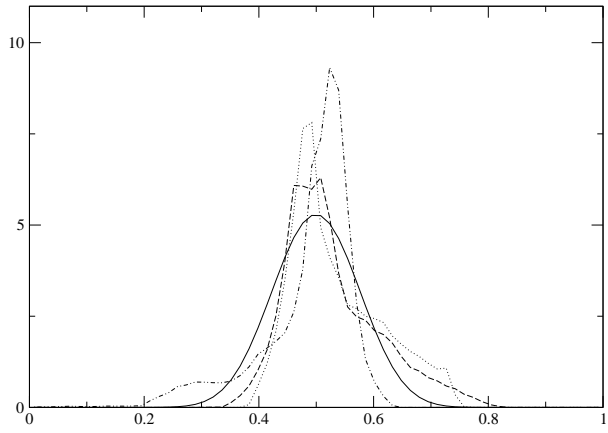}}
\caption{Comparisons between a $\beta$ law (solid line) and several
  arbitrary chosen FDFs,
of same mean ($\cb=0.5\pm 0.01$) and variance ($\vs=0.0055\pm 0.0001$)\label{unique}}
\end{center}
\end{figure}

Figure \ref{unique} shows several FDFs with extremely close means and
variances. The cubes have been chosen so that $\cb=0.5\pm 0.01$ and $\vs=0.0055\pm 0.0001$.

\section{FDF modelling}

In the framework of LES, the FDF can  be estimated either  by solving  an equation
governing its evolution\cite{gao93,pope98} or by using a model for the
FDF. The model proposed by Cook and Riley\cite{cook} uses a
presumed form for the FDFs. It has drawn much attention and
has  been the subject  of several  studies\cite{moin2000,jimenez}.
In this model, the main assumption is
that the FDF can be approximated by a $\beta$ distribution
with same mean and same variance as the real FDFs (i.e. the same first
moments). The definition of the $\beta$ distribution is as follows:
\begin{equation}\label{beta}
\beta\EP{x;\cb,\vs}=\frac{x^{a-1}(1-x)^{b-1}}{B(a,b)},
\end{equation}
with
\begin{equation}
B(a,b)=\frac{\Gamma(a)\,\Gamma(b)}{\Gamma(a+b)}=\int_0^1 x^{a-1}(1-x)^{b-1}\,\mbox{d}x
\end{equation}
$\Gamma(a)$ being the gamma function of Euler.

In order to have the right mean and variance,
$a$ and $b$ must be chosen so that
$$a=\cb\,\left(\frac{\cb\,(1-\cb)}{\vs}-1\right)~~~\mbox{and}~~~b=\frac{a}{\cb}-a.$$
The presumed form for the FDF can be used in equation
(\ref{eval}). This provides a model for $\fs$ whatever $f$ is, using
$\cb$ and $\vs$ as variables. The estimator of $\fs$ can be written
\begin{equation}
\int f(x)\,\beta\EP{x;\cb,\vs}\D x.
\end{equation}

Since the definition of the variables is crucial for the optimal estimators,
we will pay much attention to the often implicit choice of the fundamental variables.
Here $\cb$ and $\vs$ are the variables.
Hence we will denote $\pi_1=\left\{\cb,\vs\right\}$ this first set of fundamental
variables. The choice of $\beta$ distributions will be called the ``$\beta$ assumption''.

The subgrid variance $\vs$ cannot be computed using the large eddies $\cb$ only.
 A sub-model is thus necessary so that the $\beta$
assumption can be used. Let us denote $\T{.}$ a test filter of a characteristic
size twice as big as for $\F{.^{}}$. Cook and Riley assume that the subgrid variance
is proportional to the quantity
$$\alpha=\T{\cb^2}-\T{\cb}^2.$$ This estimation is used to approximate the
FDF and the whole model thus provides an estimation of $\fs$
based on the variables $\cb$ and $\alpha$ only. We will denote
$\pi_2 = \left\{\cb,\alpha\right\}$ this second set of variables.

Another sub-model has been proposed by Pierce and Moin\cite{moin98}. They assume
that $\vs$ is proportional to the modulus of the gradient of the filtered scalar, which
we will denote $\epsilon_{\cb} = (\partial_i \cb)^2$.
 In the following, we will denote
$\pi_2 = \left\{\cb,\epsilon_{\cb}\right\}$ the variables corresponding to this modelization.

\section{Validity of the $\beta$ assumption}

Our purpose is to know if there is an optimal choice for the
presumed form of the FDF. Of course this optimal choice depends on the
variables $\pi$ used for the model. The fact that there is an optimal
choice is not obvious. The optimal estimators used in the first
part are defined in the case where a scalar quantity $\gamma$ has to be
estimated using $\pi$. Here the model provides a {\em function} for each different
value of $\pi$. No relevant measure of the error can be defined in this case. But we have the relation
\begin{eqnarray}
\cond{\fs}{\pi}&=&\cond{\int f(C) \,\ds(C)\D C}{\pi}\\
&=&\int f(C)~\cond{\ds(C)}{\pi}~\mbox{d}C\label{eval2}.
\end{eqnarray}
This means that $\cond{\ds}{\pi}$ is the optimal estimator for the
approximation of the FDF using $\pi$ since when it is used in equation
(\ref{eval}), the estimator of $\fs$ which is obtained is the
optimal estimator for $\fs$ using $\pi$. In this particular case the optimal estimator concept
can therefore easily be extended. This quantity has already been
considered in the case where the variables are $\cb$ and $\vs$. It has
been called the conditional FDF\cite{tong}. It is sometimes refered to
as the FPDF\cite{pitsch}.

This optimal estimator can be computed using the following natural method:
first, grid nodes are chosen for which $\pi$ has a value very close to
an arbitrary given one, then the FDF is computed for each of these nodes, and
finally, all the FDFs are averaged to obtain the optimal
estimator.

\begin{figure}
\begin{center}
\centerline{\includegraphics[width=8cm]{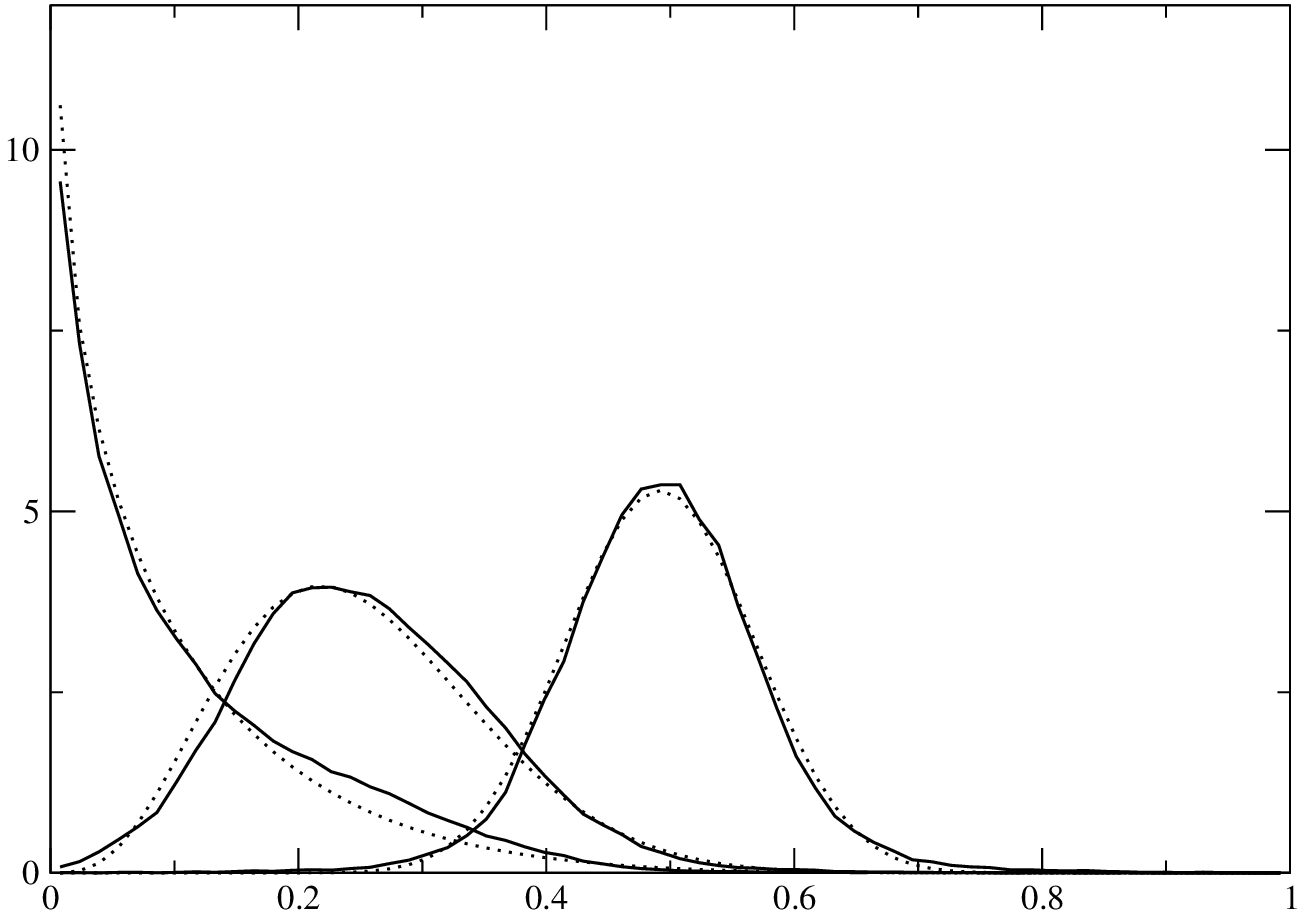}}
\caption{Comparisons between $\cond{\ds(C)}{\cb,\vs}$ and
  $\beta(C;\cb,\vs)$ for (i) $\cb=0.5$ and $\vs=0.0055$ (ii)
  $\cb=0.25$ and $\vs=0.01$ (iii) $\cb=0.1$ and $\vs=0.01$.
\label{sigma}}
\end{center}
\end{figure}

Let us consider the set of variables $\pi_1=\left\{\cb,\vs\right\}$,
which is the case when the subgrid variance is known.
Comparisons between beta distributions and
the optimal estimators are shown for three sets of values of the
variables in figure \ref{sigma}. The cubes which are selected for the
computation of the optimal estimators are chosen so that $\cb=0.5\pm 0.01$ and
$\vs = 0.0055 \pm 0.0001$ for the first comparison, $\cb=0.25\pm0.01$ and $\vs = 0.01\pm
0.001$ for the second one and $\cb=0.1\pm 0.01$ and $\vs=0.01\pm
0.001$ for the last one.

The correspondence between the $\beta$ distributions and the optimal
estimators is excellent. The $\beta$ distributions can thus be considered
as a very appropriate presumed form for the FDF, as long as $\vs$ is
known. We must point out that this does not
mean that the FDFs are actually $\beta$ distributions\cite{cook}, as shown figure \ref{unique}.
We agree \cite{cook} that the $\beta$ assumption seems to be
appropriate for any subvolume. We think that a mathematical property of
$\beta$ distributions could explain these results but we were not able
to find it. Here, the size filter is about four
grid nodes and has been chosen so that all values of $\cb$, $\vs$ (or
$\alpha$) are well represented in the statistical sampling process.

\begin{figure}
\begin{center}
\centerline{\includegraphics[width=8cm]{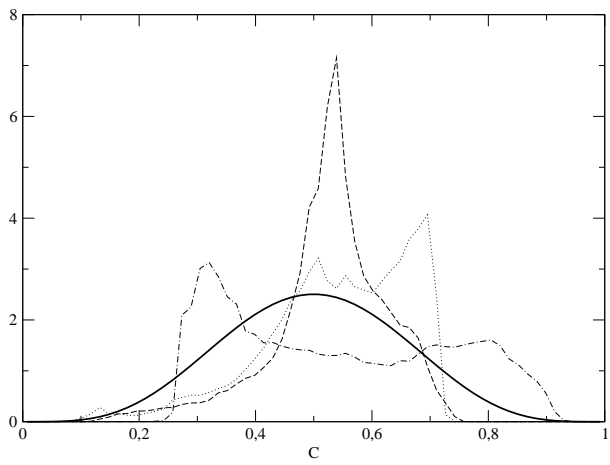}}
\caption{Comparisons between a $\beta$ law (solid line) and several
  arbitrary chosen FDFs, for $\cb=0.5\pm 0.01$
  and $\alpha=0.0125\pm 0.0005$.\label{unique-alpha}}
\end{center}
\end{figure}

Let us consider the case when the variables of the model are $\cb$
and $\alpha$. Figure \ref{unique-alpha} shows examples of FDFs for cubes which present
the same values of $\cb$ and $\alpha$. When compared to figure \ref{unique}, it is observed
that there are much larger differences between the FDFs. This is due
to the fact that for given $\cb$ and $\alpha$, different values of $\vs$ are observed.
This is what we will call the {\em subgrid variance dispersion}. For a given $\pi$
the subgrid variance dispersion can be quantified by the irreducible error
made when estimating $\vs$ using $\pi$. When for instance this error is small,
the subgrid variance of cubes with very close $\pi$ values will be
very close to $\cond{\vs}{\pi}$.

\begin{figure}
\begin{center}
\centerline{\includegraphics[width=8cm]{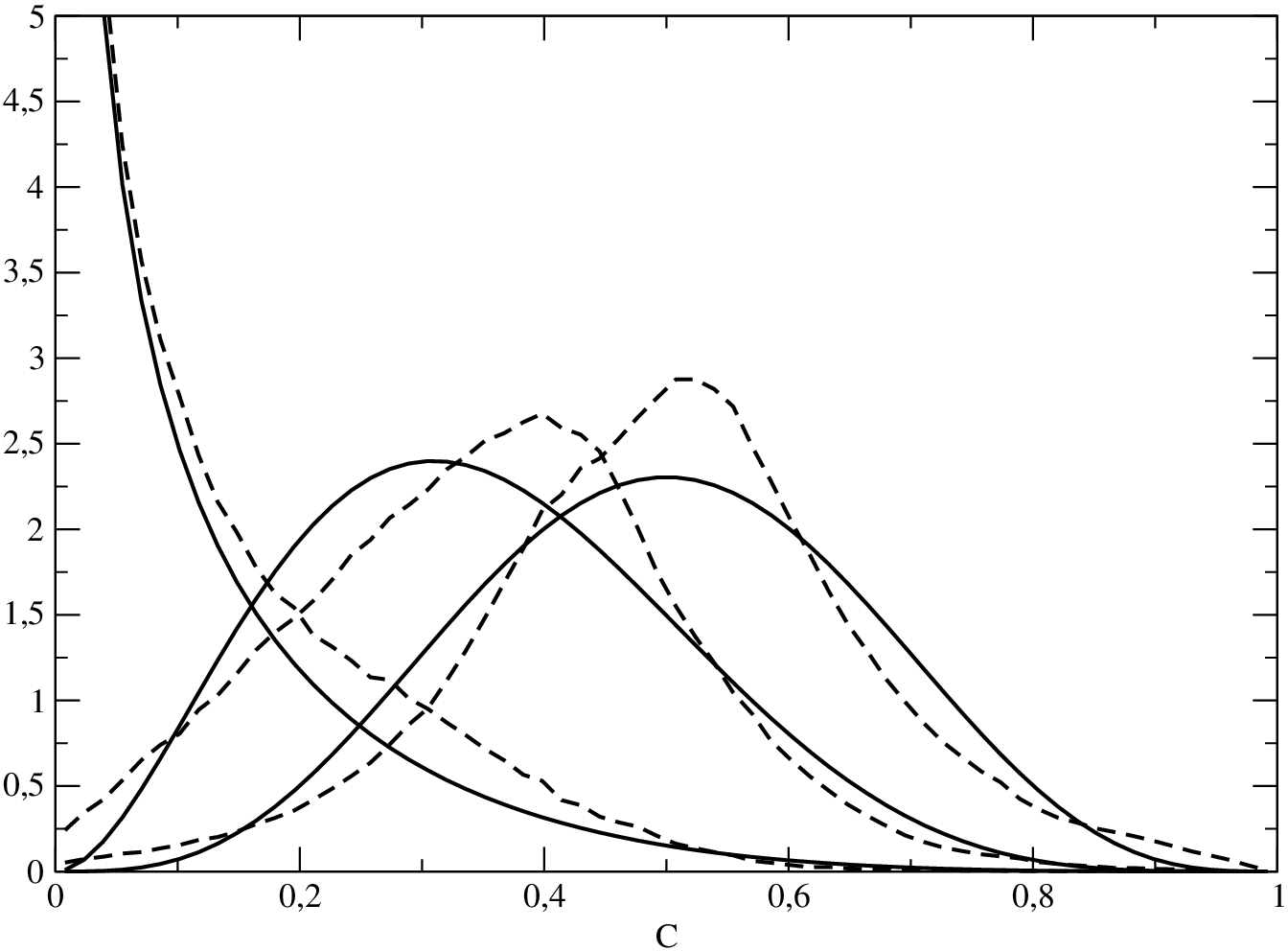}}
\caption{Comparisons between $\cond{\ds(C)}{\cb,\alpha}$ and
$\beta(C;\cb,\cond{\vs}{\cb,\alpha})$\label{alpha}}
\end{center}
\end{figure}

Figure \ref{alpha}, the optimal estimators are compared to a $\beta$ distribution whose variance is
$\cond{\vs}{\cb,\alpha}$ (the average of $\vs$ for all the
FDFs). The cubes which are selected for the
computation of the optimal estimators are chosen so that
$\cb=0.5\pm 0.01$ and $\alpha=0.0125\pm 0.0005$ for the first case,
$\cb=0.355\pm 0.005$ and $\alpha = 0.01\pm 0.001$ for the second case, and for the last one $\cb =
0.1\pm 0.01$ and $\alpha = 0.001\pm 0.001$.

In this case, the $\beta$ distributions are not very close to the
optimal estimators. This is due to the variance dispersion.
This means that there is a better presumed form when $\alpha$ is
used - closer to the optimal estimator. However, this form would be
relevant for $\pi=\{\cb,\alpha\}$ only.

When choosing a set $\pi$ with
less subgrid variance dispersion, the form of the optimal
estimator must tend towards the form of the optimal estimator when $\vs$
is known (when the dispersion is null). Hence, we conclude that the
$\beta$ assumption is better when the subgrid variance dispersion
is small.

Now that we have established that, when $\vs$ is unknown, $\beta$ distributions are not
as clearly appropriate as when the subgrid variance is known, the question is: is
the error due to the difference between the optimal estimator of the
FDF and the $\beta$ distribution a significant one ?

Using optimal estimators, it is possible to compute the supplementary
error brought by the $\beta$ assumption {\bf alone} for the estimation of
$\fs$. This must be done for a given $f$.

Let us consider the following estimator for $\fs$ using $\pi$:
\begin{equation}
g_f(\pi)=\int  f(C) ~ \beta\EP{C ; \F{c},\cond{\vs}{\pi}}
~\mbox{d}C.\label{e:est}
\end{equation}
It is obtained by replacing $\ds$ in (\ref{eval}) by a $\beta$
distribution. The variance of the $\beta$ distribution is the optimal
estimator of $\vs$ using $\pi$. The error made by this estimator can
be compared to the irreducible error made by $\cond{\fs}{\pi}$. The
supplementary error made by the estimator (\ref{e:est}) reflects the
fact that $\beta$ distributions are not exactly the optimal estimators as
shown figure \ref{alpha} (or figure \ref{sigma} even if the
difference is slight).

We will now present results that have been obtained using
histograms. All the quadratic errors have been {\em normalized by the variance} of the quantity
which is estimated. This allows to compare two errors even if the
quantity to estimate is not the same.

The results concerning the estimation of $\vs$ are presented in table \ref{t1} concerning the estimation
of $\vs$. The irreducible error that is computed here reflects the
subgrid variance dispersion. This dispersion is smaller when the
variables are $\cb$  and $\epsilon_{\cb}$ than when the variables are
$\cb$ and $\alpha$.

Since the error brought by the $\beta$ assumption must be computed for a
given $f$ we have chosen to do so for several $\beta$
distributions with different means and variances. This is a very plausible
choice\cite{moin98}. The comparison between the optimal estimators and
the estimator $g_f(\pi)$ is presented in table \ref{tableau}.
Column one is the error made by the optimal estimators, i.e. the
irreducible error. Column two is the error made by $g_f(\pi)$, i.e.
when using the $\beta$ assumption.

It can be observed that the supplementary error due to the $\beta$
assumption is always much smaller than the irreducible error. In most
cases, it is one order of magnitude smaller. This confirms the
previous results on the optimal estimators of the FDF.

The fact that the $\beta$ distributions are not very close to the
optimal estimators of the FDFs has not a measurable incidence on the
supplementary error. The latter can often be neglected and there is
no need to search for a more relevant presumed form for the FDF.

Since the irreducible errors are normalized, they can be compared. The
conclusion is that $\vs$ is a quantity which is very difficult to
estimate. The quantity $\fs$ is in general easier to estimate. In
addition, the better $\vs$ is estimated using a set of variables,
the better the $\fs$ are estimated. For a few variables,
the relative relevancy of a set does not depend on $f$.

\begin{table}
\begin{center}
\begin{tabular}{l c}
Estimator  & Normalized error \\
\hline \hline
$\cond{\vs}{\cb,\epsilon_{\F{c}}}$  & 0.215  \\ \hline
$\cond{\vs}{\cb,\alpha}$            & 0.259  \\ \hline
$\vs = \kappa\, \alpha$             & 0.359  \\ \hline
\end{tabular}
\caption{Normalized quadratic errors for several optimal estimators of
  $\vs$ (the optimal estimators are computed using the histogram
  technique). The constant $\kappa$ is chosen so that the error made by the Cook
  and Riley estimator of $\vs$ is minimized.\label{t1}}
\end{center}
\end{table}

\begin{table}
\begin{center}
\begin{tabular}{l c c c }
Variables ($\pi$) & Error of $\cond{\fs}{\pi}$ & Error of $g_f(\pi)$ \\
\hline \hline
\multicolumn{2}{l}{Mean 0.35, variance 0.01} \\ \hline
$\cb$, $\vs$             &  0.043  &  0.051 \\ \hline
$\cb$,$\epsilon_{\F{c}}$ &  0.095  &  0.099 \\ \hline
$\cb$,$\alpha$           &  0.136  &  0.140 \\
\hline \hline
\multicolumn{2}{l}{Mean 0.15, variance 0.01} \\ \hline
$\cb$, $\vs$             &  0.031  &  0.042 \\ \hline
$\cb$,$\epsilon_{\F{c}}$ &  0.055  &  0.070 \\ \hline
$\cb$,$\alpha$           &  0.072  &  0.091 \\
\hline \hline
\multicolumn{2}{l}{Mean 0.5, variance 0.01} \\ \hline
$\cb$, $\vs$             &  0.041  &  0.063 \\ \hline
$\cb$,$\epsilon_{\F{c}}$ &  0.092  &  0.107 \\ \hline
$\cb$,$\alpha$           &  0.130  &  0.146 \\
\hline \hline
\multicolumn{2}{l}{Mean 0.5, variance 0.036} \\ \hline
$\cb$, $\vs$             &  0.011  &  0.013 \\ \hline
$\cb$,$\epsilon_{\F{c}}$ &  0.064  &  0.066 \\ \hline
$\cb$,$\alpha$           &  0.079  &  0.081 \\ \hline
\end{tabular}
\caption{Normalized quadratic errors of different estimators for the
  estimation of $\fs$. The mean and the variance of the different $f$
  that have been chosen are specified.\label{tableau}}
\end{center}
\end{table}

\section{Neural networks}\label{s:nn}

As already underlined\cite{cook}, the estimation of the
subgrid variance $\vs$ is the main problem in the presumed FDF
approach. Rather complex models have been proposed\cite{moin2000}
using dynamic approach and many new variables. Table \ref{tab:neurones} shows the irreducible
error for different sets of variables computed using neural
networks. One of these variables ($\cb$) has often been neglected
in the different models proposed. The results shown here suggest this
is a mistake. On the contrary, the results show that the dissipation $\epsilon$ does not bring much
information about $\vs$.

\begin{table}
\begin{center}
\begin{tabular}{c c}
\mbox{\it Variables} & \mbox{\it Irreducible error} \\\hline \hline
$\cb,\alpha$ & 0,259 \\
$\cb,\dcb$ & 0,215 \\
$\cb,\alpha,\dcb$ & 0,192\\
$\cb,\alpha,\epsilon$ & 0,258\\
$\alpha, \dcb,\epsilon_{\T{\cb}},\T{\dcb}$ & 0,173 \\
$\cb,\alpha,\dcb,\epsilon_{\T{\cb}}$ & 0,158 \\
$\cb,\alpha,\dcb,\T{\dcb}$ & 0,152 \\
$\cb,\alpha,\dcb,\epsilon_{\T{\cb}},\T{\dcb}$ & 0,137\\
\hline
\end{tabular}
\caption{Irreducible error linked to different parameter sets. 
$64^3$ examples are used for evaluating the weights of the neural
networks and $(128^3-64^3)$ examples are used for estimating the
irreducible error.\label{tab:neurones}}
\end{center}
\end{table}

In addition, they show that the more variables are included in $\pi$,
the better the estimation of the subgrid quantities - with errors much
smaller than with usual models. Neural network
can thus be considered as a new way towards {\it optimal LES} in the sense
of Langford and Moser\cite{langford}. We think that they could be used
directly for the modelling of unknown terms in LES and that they
present some advantages: they do not need much data to estimate
the conditional expectations correctly and they can reproduce highly non linear
behaviours. It is possible to take physics directly into account when choosing
the variables (galilean invariance\cite{langford}, scale invariance
or dimensional analysis arguments).

\section{Conclusion}

In agreement with Langford and Moser\cite{langford}, we have 
used and explored the idea that in the framework of LES,
once a set of variables has been chosen to estimate a given subgrid term,
there is only one optimal estimator. Some of
the properties of optimal estimators that are of interest for LES
have been presented throughout this article. 
We have shown how optimal estimators could be
computed using simple techniques if the number of variables is
small, or neural networks if the number of variables is greater
than 3.

As an illustration, the concept of optimal estimator was applied
to the Cook and Riley subgrid model\cite{cook} for the scalar
fluctuation, a model which is widely used and whose basis has
already retained much attention in the literature. The concept of
optimal estimator has been extended to the approximation of FDF
and the main conclusion is that the $\beta$ assumption is very
appropriate. When the error directly associated with the $\beta$
assumption for the estimation of a given quantity is compared with
the irreducible error, it is found to be small. The largest
error are associated with the estimation of the subgrid variance
of the scalar fluctuations. That is the very point on which the
Cook and Riley model needs to be improved.

In the paper we did not attempt to propose any new practical model
for this scalar variance, but we used neural networks to see how
closely the subgrid variance could be estimated. The error made by
the optimal estimator is smaller when the number of relevant
variables is increased.

The optimal estimation technique provides a way of assesssing which
set of parameters will potentially lead to the most accurate
subgrid model. It does not provide any information on how to
specify the formulation of the model, but simply indicates if
efforts for building a model with a given set of parameters are
likely to be fruitful.

As the number of retained parameters is increased, it can become
more and more difficult to propose a model formulation on the
ground of physical reasoning, and this might suggest using the
optimal estimator directly instead of a model. Then, neural
networks would be used directly - instead of a modelization.
This ``NN guided LES'' could constitute an {\it optimal
LES} as defined by\cite{langford,volker}. We did not perform
such a simulation in the present paper. This could be the
subject of a future study (following the path explored by 
\cite{volker}).

This paper is an illustration of the relevance of optimal
estimation techniques for the problem of subgrid modeling for LES.
These techniques allow a thorough exploration of the behaviours of
models in LES indicating the points which have to be improved
in the models.

\section*{Acknowledgements}

Olivier Teytaud is supported in part by the Pascal Network of Excellence.
The authors would like to thank M.F. Moreau for a careful reading of
the paper.

\appendix

\section{Orthogonality relation}
Let us give a short demonstration of (\ref{e:orth}). The quadratic error
made by an estimator $g(\pi)$ can be developed:

\begin{eqnarray*}
\erreur{\gamma}{g(\pi)} &=& \erreur{\gamma}{\cond{\gamma}{\pi}}+\erreur{\cond{\gamma}{\pi}}{g(\pi)}\\
& & + 2 \moy{(\gamma-\cond{\gamma}{\pi})\,(\cond{\gamma}{\pi}-g(\pi))}
\end{eqnarray*}
Now, we have to show that the last term is null. It can be written
\begin{eqnarray*}
& &2\,\int (\gamma-\cond{\gamma}{\pi})\,(\cond{\gamma}{\pi}-g(\pi))
\,p(\gamma,\pi) \D\gamma \D\pi \\
&=& 2\,\int (\cond{\gamma}{\pi}-g(\pi))\,p(\pi) \left( \int
(\gamma-\cond{\gamma}{\pi})\,p(\gamma|\pi) \D\gamma\right)\D\pi
\end{eqnarray*}
Since $\cond{\gamma}{\pi}= \int \gamma p(\gamma|\pi)\D\gamma$, the
previous quantity is null and the orthogonality relation holds.

\section{Successive conditioning}

Let $g(\pi)$ be an estimator of $\gamma$ using $\pi$. We will note
$p(g(\pi)=g)$ or $p(g)$ the PDF that $g(\pi)=g$. Then we have
\begin{eqnarray*}
\cond{\gamma}{g(\pi)=g}\,p(g(\pi)=g) &=& \int \gamma \,p(\gamma,g)
\D\gamma\\
&=&\int \gamma \,p(\gamma,\pi,g)\D\gamma\D\pi\\
&=&\int \gamma \,p(\gamma,\pi)\,\delta(g(\pi)-g) \D\gamma\D\pi\\
&=&\int \left(\int \gamma \,p(\gamma|\pi)\D\gamma\right)\,p(\pi)\,\delta(g(\pi)-g)
\D\pi\\
&=&\int \cond{\gamma}{\pi} \,p(\pi)\,\delta(g(\pi)-g)
\D\pi.
\end{eqnarray*}
And if $g(\pi)=\cond{\gamma}{\pi}$ then
\begin{eqnarray*}
\cond{\gamma}{\cond{\gamma}{\pi}=g}\,p(g) &=& \int
\cond{\gamma}{\pi}\,p(\pi)
\,\delta(\cond{\gamma}{\pi}-g)\D\pi\\
&=&g\,\int p(\pi) \,\delta(\cond{\gamma}{\pi}-g)\D\pi\\
&=& g \int p(\pi,g)\D\pi\\
&=& g\,p(g)
\end{eqnarray*}
Hence,
$\cond{\gamma}{g(\pi)=g} = g$, which can be written
$$\cond{\gamma}{\cond{\gamma}{\pi}}= \cond{\gamma}{\pi}.$$

\section{Filtering}

The filtering can be written
$$\cb(\V{r}) = \int G(\V{r}-\V{x})\,c(\V{x})\D\V{x}$$
or $\cb=G*c$. Let us assume that we have $0\leq c\leq 1$ and that we want
\begin{equation}0\leq\cb\leq 1.\label{e:bornes}\end{equation}
If $G(\V{x})\geq 0\;\forall \V{x}$ we can write
$$0\leq\cb\leq \F{1}.$$
If $G$ is not positive, the inequality can be false. Thus it is
necessary that $G$ should be positive. In order to have
(\ref{e:bornes}) verified, $G$ must have the property $\F{1}=1$,
which can be written $\int G = 1$.

Let us now define $V_D(C)=\F{H(C-c(\V{x}))}$. The physical meaning of
this quantity is not as clear as in the case of the box
filter. Anyway, the FDF is still the derivative of $V_D$. We can write
$$V_D(a)-V_D(b) =\F{H(a-c)-H(b-c)}.$$
If $a\geq b$ then $H(a-c(\V{x})-H(b-c(\V{x}))\geq 0$. {\it If the filter is
positive}, then $V_D(a)\geq V_D(b)$ and $V_D$ is a growing function.
If $G$ is not positive, this is not granted. Since the FDF is the
derivative of $V_D$, it is positive only if the filter is positive.
Finally, we have $V_D(1)=\F{1}=\int \ds$, so that the FDF is
normalized only if $\int G =1$.

The filters are generally chosen so that $\int G =1$. The previous
arguments show that they should be chosen positive, too. The box filter has
been chosen for historical\cite{cook,moin2000} and clarity reasons.
But it is an invertible filter, whereas the filters that should be
considered in the framework of LES should be non-invertible\cite{langford}.
All the non-invertible filters that have been proposed are not positive.
We propose here a new filter that could be more relevant. It is
defined by
\begin{equation}
G(\V{x}) = \prod_{i=1}^{3} \frac{2}{\pi\,\ell}\; \mbox{sinc}^2 \frac{x}{\ell}.
\end{equation}
This filter is non-invertible (since the Fourier Transform of $\mbox{sinc}^2$ is a
triangle function), normalized and positive. We think that the choice
of a particular filter has
not much importance while the error made by the estimators is
 large. For very small errors, a difference could probably be
found between the error made by the estimators when using an invertible
filter and the error when using a non-invertible filter.
This could probably be seen using neural networks when computing the
irreducible error.

\end{document}